# DNN AND CNN WITH WEIGHTED AND MULTI-TASK LOSS FUNCTIONS FOR AUDIO EVENT DETECTION

*Huy Phan*\*, *Martin Krawczyk-Becker*†, *Timo Gerkmann*†, and *Alfred Mertins*\*

\* University of Lübeck, Institute for Signal Processing, Lübeck, Germany
† University of Hamburg, Department of Informatics, Hamburg, Germany
{phan,mertins}@isip.uni-luebeck.de, {krawczyk,gerkmann}@informatik.uni-hamburg.de

## ABSTRACT

This report presents our audio event detection system submitted for Task 2, "Detection of rare sound events", of DCASE 2017 challenge [1]. The proposed system is based on convolutional neural networks (CNNs) and deep neural networks (DNNs) coupled with novel weighted and multi-task loss functions and state-of-the-art phase-aware signal enhancement. The loss functions are tailored for audio event detection in audio streams. The weighted loss is designed to tackle the common issue of imbalanced data in background/foreground classification while the multi-task loss enables the networks to simultaneously model the class distribution and the temporal structures of the target events for recognition. Our proposed systems significantly outperform the challenge baseline, improving F-score from 72.7% to 90.0% and reducing detection error rate from 0.53 to 0.18 on average on the development data. On the evaluation data, our submission obtains an average F1-score of 88.3% and an error rate of 0.22 which are significantly better than those obtained by the DCASE baseline (i.e. an F1-score of 64.1% and an error rate of 0.64).

## 1. INTRODUCTION

There is an ongoing methodological trend in computational auditory scene analysis (CASA), shifting from conventional methods to modern deep learning techniques [2, 3, 4, 5, 6]. However, most of the works have focused on the aspect of network architectures which have been usually adapted from those successful in related fields, such as computer vision and speech recognition. Little attention has been paid to loss functions of the networks. Although the common loss functions, such as the cross-entropy loss for classification and the $\ell_2$-distance loss for regression, work for general settings, it is arguable that the loss functions should be tailored for a particular task at hand.

In this work, we propose two such tailored loss functions, namely *weighted loss* and *multi-task loss*, coupled with common deep network architectures to tackle the well-known issues of audio event detection (AED). The weighted loss can be used to explicitly weight penalties for two types of errors (i.e. false negative and false positive errors) in a binary classification problem. This loss is, therefore, useful for imbalanced background/foreground classification in AED in which the foreground samples are more valuable than the numerous background samples and should be penalized stronger if misclassified. The multi-task loss, however, is proposed to suit classification of target events. As audio events possess inherent temporal structures, modeling them has been shown important for recognition [7, 8, 9] and detection [10, 11]. The multi-task loss is designed to allow a network to model both event class distribution (as a classification task) and event temporal structures (as a regression task for event onset and offset estimation) at the same time. By doing this, the network is forced to cope with a more complex problem rather than the simple classification one. As a result, the network is implicitly regularized, leading to improvements of its generalization capability. Obviously, an inference step like the one in [10, 12] can be further performed for real-time early event detection in continuous streams.

In this work, we study the coupling of the proposed loss functions with both deep neural networks (DNNs) and convolutional neural networks (CNNs) for audio event detection. Experimental results conducted on the development data of the DCASE 2017 challenge show that the proposed systems significantly outperform the challenge's baseline system.

## 2. THE PROPOSED DETECTION SYSTEM

The overall pipeline of the proposed detection system is illustrated in Figure 1. The audio signals are firstly preprocessed for signal enhancement (cf. Section 2.1). The preprocessed signals are then decomposed into small frames and frame-wise feature extraction is performed. We commonly employ log Gammatone spectral coefficients [13] for both DNN-based and CNN-based systems. However, we tailored the feature extraction strategies to produce suitable inputs for individual network types (cf. Section 2.2).

Although Task 2 of the challenge is set up to evaluate detection of three categories (baby cry, glass break, and gun shot) separately, our proposed systems are multi-class, aiming at detecting all the three target categories at once. By doing this, we avoid optimizing different systems for individual categories. The proposed systems accomplish the detection goal in two steps: background rejection and event classification. The former uses a binary classifier to filter out background frames and lets only foreground frames go through. Subsequently, the latter employs a multi-class classifier to distinguish the frames identified as foreground into three target categories. We investigate both DNNs and CNNs for classification.

Two networks (i.e. two DNNs for the DNN-based system and two CNNs for the CNN-based system) are employed, one for background rejection and the other for subsequent event classification. Both networks share a similar architecture, except for the dropout probability, the output layers, and the loss functions which are task-dependent. The DNN architecture and its associated parameters are shown in Figure 2 and Table 1, respectively, while those of the CNN are shown in Figure 3 and Table 2. The task-dependent output layers and loss functions will be described in more detail in Sections 2.3 and 2.4.

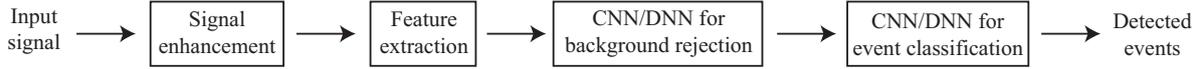

Figure 1: The overall pipeline of the proposed audio event detection system.

## 2.1. Phase-aware signal enhancement

For all three categories, baby cry, glass break, and gun shot, short-time discrete Fourier transform (STFT) domain signal enhancement was employed to reduce acoustic noise in the recordings. The STFT segments had a length of 32 ms with consecutive segments overlapping by 50 %. For analysis and synthesis, a square-root Hann window was used. The STFT magnitudes of the clean signals were estimated from the noisy signals according to [14], with its parameters set to $\mu^{[14]} = \beta^{[14]} = 0.5$, and combined with the noisy phase for the reconstruction of the enhanced time domain signal. The magnitude estimation in [14] relies on the power spectral densities (PSDs) of noise and speech as well as estimates of the clean STFT phase. The speech PSD was estimated via [15] and the noise PSD via temporal cepstrum smoothing [16, 17]. Estimates of the clean STFT phase were obtained according to [18], which in turn relies on estimates of the fundamental frequency of the desired sound. Accordingly, [18] provides estimates of the clean phase only for sounds for which a fundamental frequency is defined, i.e. harmonic sounds such as baby cries. Harmonic sounds and their fundamental frequency were found using the noise robust fundamental frequency estimator PEFAC [19]. To focus on baby cries, we limited the search range of PEFAC to frequencies between 300 Hz and 750 Hz, which covers the relatively high fundamental frequency of most baby cries while excluding lower frequencies that are found in adult speech. As proposed in [14], for all non-voiced sounds we employed the phase-blind spectral magnitude estimator [20], which does not need any clean phase estimate.

Finally, to avoid undesired distortions of the desired signal, we limited the maximum attenuation that can be applied to each STFT time-frequency point to 12 dB.

## 2.2. Feature extraction

The feature extraction step was accomplished differently for the DNN- and CNN-based systems.

For the former, an audio signal was decomposed into frames of length 100 ms with a hop size of 20 ms. 64 log Gammatone spectral coefficients [13] in the frequency range of 50 Hz to 22050 Hz were then extracted for each frame. In addition, we considered a context of five frames for classification purpose. The feature vector for a context window was formed by simply concatenating feature vectors of its five constituent frames.

For the latter, we opted a frame size of 40 ms and a hop size of 20 ms for signal decomposition. A feature set of 64 log Gammatone spectral coefficients was then calculated for each frame as in the DNN case. In addition, delta and acceleration coefficients were also calculated using a window length of nine frames. Eventually, 64 consecutive frames are combined into a $64 \times 64 \times 3$ image which was used as input for the CNNs.

## 2.3. Background rejection with weighted loss

In general, for audio event detection in continuous streams, the number of background frames is significantly larger than for foreground ones. This leads to a skewed classification problem with a dominance of the background samples. The skewness is even more severe in case of the "Detection of rare events" task. To remedy this skewness issue, in combination with data resampling, we propose a *weighted loss* function to train the networks for background rejection.

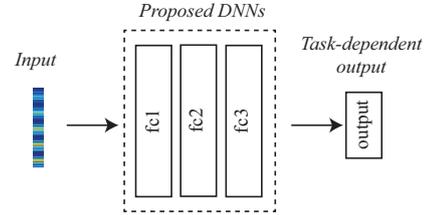

Figure 2: The proposed DNN architecture.

Table 1: The parameters of the DNN architecture. A dropout probability of 0.5 and 0.2 is used for background rejection and event classification, respectively.

| Layer | Size | Activation | Dropout |
|-------|------|------------|---------|
| fc1 | 512 | ReLU | 0.5/0.2 |
| fc2 | 256 | ReLU | 0.5/0.2 |
| fc3 | 512 | ReLU | 0.5/0.2 |

Firstly, the background samples were downsampled by a factor of 5. Furthermore, the set of foreground samples was upsampled by an integer factor to make its size approximately equal to the background set. Let us denote a training set of $N$ training examples as $\{(\mathbf{x}_1, \mathbf{y}_1), \ldots, (\mathbf{x}_N, \mathbf{y}_N)\}$ where $\mathbf{x}$ denotes a one-dimensional feature vector (in case of DNN) or a three-dimensional image (in case of CNN). $\mathbf{y} \in \{0, 1\}^C$ denotes a binary one-hot encoding vector with $C = 2$ in this case.

Typically, for a classification task, a network will be trained to minimize the cross-entropy loss

$$E(\boldsymbol{\theta}) = -\frac{1}{N} \sum_{n=1}^{N} \mathbf{y}_n \log \left( \hat{\mathbf{y}}_n(\mathbf{x}_n, \boldsymbol{\theta}) \right) + \frac{\lambda}{2} \|\boldsymbol{\theta}\|_2^2, \quad (1)$$

where $\boldsymbol{\theta}$ denotes the network's trainable parameters and the hyper-parameter $\lambda$ is used to trade-off the error term and the $\ell_2$-norm regularization term. The predicted posterior probability $\hat{\mathbf{y}}(\mathbf{x}, \boldsymbol{\theta})$ is obtained by applying the softmax function on the network output layer. However, this loss penalizes different classification errors equally. In contrast, our proposed weighted loss enables to penalize individual classification errors differently. The weighted loss reads

$$E_{\mathrm{w}}(\boldsymbol{\theta}) = -\frac{1}{N} \left( \lambda_{\mathrm{fg}} \sum_{n=1}^{N} \mathbb{I}_{\mathrm{fg}}(\mathbf{x}_n) \mathbf{y}_n \log \left( \hat{\mathbf{y}}_n(\mathbf{x}_n, \boldsymbol{\theta}) \right) \right.$$
$$\left. + \lambda_{\mathrm{bg}} \sum_{n=1}^{N} \mathbb{I}_{\mathrm{bg}}(\mathbf{x}_n) \mathbf{y}_n \log \left( \hat{\mathbf{y}}_n(\mathbf{x}_n, \boldsymbol{\theta}) \right) \right) + \frac{\lambda}{2} \|\boldsymbol{\theta}\|_2^2, \quad (2)$$

where $\mathbb{I}_{\mathrm{fg}}(\mathbf{x})$ and $\mathbb{I}_{\mathrm{bg}}(\mathbf{x})$ are indicator functions which specify whether the sample $\mathbf{x}$ is foreground or background, respectively. $\lambda_{\mathrm{fg}}$ and $\lambda_{\mathrm{bg}}$ are *penalization weights* for false negative errors (i.e. a foreground sample is misclassified as background) and false positive errors (i.e. a background sample is misclassified as foreground), respectively. Since foreground samples are more valuable than background ones in the skewed classification problem at hand, we penalize false negative errors more than false positive ones (cf. Section 3.2).

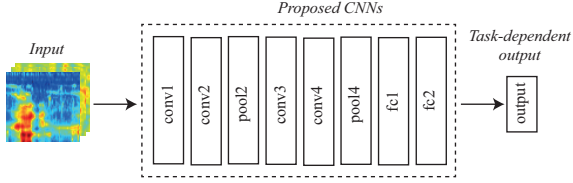

Figure 3: The proposed CNN architecture.

Table 2: The parameters of the CNN architecture. The number of feature maps and the dropout probability are set to 64 and 0.5, respectively, for background rejection while they are set to 128 and 0.2, respectively, for event classification.

| Layer | Size | #Fmap | Activation | Dropout |
|-------|------|-------|------------|---------|
| conv1 | 3 × 3 | 64/128 | ReLU | - |
| conv2 | 3 × 3 | 64/128 | ReLU | - |
| maxpool2 | 2 × 1 | - | - | 0.5/0.2 |
| conv3 | 3 × 3 | 64/128 | ReLU | - |
| conv4 | 3 × 3 | 64/128 | ReLU | - |
| maxpool4 | 2 × 2 | - | - | 0.5/0.2 |
| fc1 | 1024 | - | ReLU | 0.5/0.2 |
| fc2 | 1024 | - | ReLU | 0.5/0.2 |

### 2.4. Event classification with multi-task loss

Beyond a simple event classification, we enforce the networks to jointly model the class distribution for event classification and the event temporal structures for onset and offset distance estimation similar to [21]. The proposed *multi-task loss* is specialized for this purpose. Multi-task modeling can be interpreted as implicit regularization which is expected to improve generalization of a network [22, 23, 24]. Furthermore, although it has not been done in this work, the inference step can be performed similarly to [10, 12] for early event detection in audio streams.

Similar to [10, 12], in addition to the one-hot encoding vector $\mathbf{y} \in \{0, 1\}^C$ ($C = 3$ here), we associated a sample $\mathbf{x}$ with a distance vector $\mathbf{d} = (d_{\mathrm{on}}, d_{\mathrm{off}}) \in \mathbb{R}^2$. $d_{\mathrm{on}}$ and $d_{\mathrm{off}}$ denote the distances from the center frame of $\mathbf{x}$ to the corresponding event onset and offset. The onset and offset distances were normalized to $[0, 1]$.

The output layer of a multi-task network (i.e. a DNN or a CNN) consists of two variables: $\bar{\mathbf{y}} = (\bar{y}_1, \bar{y}_2, \ldots, \bar{y}_C)$ and $\bar{\mathbf{d}} = (\bar{d}_{\mathrm{on}}, \bar{d}_{\mathrm{off}})$ as illustrated in Figure 4. The network predictions for class posterior probability $\hat{\mathbf{y}} = (\hat{y}_1, \hat{y}_2, \ldots, \hat{y}_C)$ and distance vector $\hat{\mathbf{d}} = (\hat{d}_{\mathrm{on}}, \hat{d}_{\mathrm{off}})$ are then obtained by:

$$\hat{\mathbf{y}} = \mathrm{softmax}(\bar{\mathbf{y}}), \tag{3}$$

$$\hat{\mathbf{d}} = \mathrm{sigmoid}(\bar{\mathbf{d}}). \tag{4}$$

Given a training set $\{(\mathbf{x}_1, \mathbf{y}_1, \mathbf{d}_1), \ldots, (\mathbf{x}_N, \mathbf{y}_N, \mathbf{d}_N)\}$ of $N$ samples, the network is trained to minimize the following *multi-task loss* function:

$$\begin{aligned} E_{\mathrm{mt}}(\boldsymbol{\theta}) = & \ \lambda_{\mathrm{class}} E_{\mathrm{class}}(\boldsymbol{\theta}) + \lambda_{\mathrm{dist}} E_{\mathrm{dist}}(\boldsymbol{\theta}) \\ & + \lambda_{\mathrm{conf}} E_{\mathrm{conf}}(\boldsymbol{\theta}) + \frac{\lambda}{2} \|\boldsymbol{\theta}\|_2^2, \end{aligned} \tag{5}$$

where

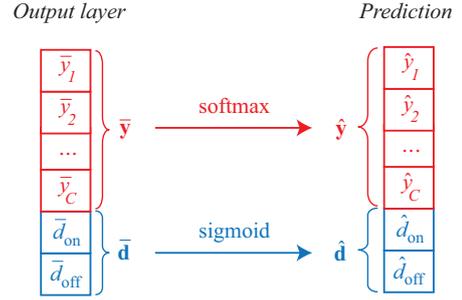

Figure 4: The output layer and the prediction of a multi-task network (i.e. a DNN or a CNN).

$$E_{\mathrm{class}}(\boldsymbol{\theta}) = -\frac{1}{N} \sum_{n=1}^{N} \mathbf{y}_n \log \left( \hat{\mathbf{y}}_n(\mathbf{x}_n, \boldsymbol{\theta}) \right), \tag{6}$$

$$E_{\mathrm{dist}}(\boldsymbol{\theta}) = -\frac{1}{N} \sum_{n=1}^{N} \left\| \mathbf{d} - \hat{\mathbf{d}}_n(\mathbf{x}_n, \boldsymbol{\theta}) \right\|_2^2, \tag{7}$$

$$E_{\mathrm{conf}}(\boldsymbol{\theta}) = -\frac{1}{N} \sum_{n=1}^{N} \left\| \mathbf{y}_n - \hat{\mathbf{y}}_n \frac{I \left( \mathbf{d}_n, \hat{\mathbf{d}}_n(\mathbf{x}_n, \boldsymbol{\theta}) \right)}{U \left( \mathbf{d}_n, \hat{\mathbf{d}}_n(\mathbf{x}_n, \boldsymbol{\theta}) \right)} \right\|_2^2. \tag{8}$$

$E_{\mathrm{class}}(\boldsymbol{\theta})$, $E_{\mathrm{class}}(\boldsymbol{\theta})$, and $E_{\mathrm{conf}}(\boldsymbol{\theta})$ in above equations are so-called *class loss*, *distance loss*, and *confidence loss*, respectively. The terms $\lambda_{\mathrm{class}}$, $\lambda_{\mathrm{dist}}$, and $\lambda_{\mathrm{conf}}$ represent the weighting coefficients for three corresponding loss types. The *class loss* complies with the common cross-entropy loss to penalize classification errors whereas the *distance loss* penalizes distance estimation errors. Furthermore, the *confidence loss* penalizes both classification errors and distance estimation errors. The functions $I \left( \mathbf{d}, \hat{\mathbf{d}} \right)$ and $U \left( \mathbf{d}, \hat{\mathbf{d}} \right)$ in (8) calculate the intersection and the union of the ground-truth event boundary and the predicted one, given by:

$$I \left( \mathbf{d}, \hat{\mathbf{d}} \right) = \min \left( d_{\mathrm{on}}, \hat{d}_{\mathrm{on}} \right) + \min \left( d_{\mathrm{off}}, \hat{d}_{\mathrm{off}} \right), \tag{9}$$

$$U \left( \mathbf{d}, \hat{\mathbf{d}} \right) = \max \left( d_{\mathrm{on}}, \hat{d}_{\mathrm{on}} \right) + \max \left( d_{\mathrm{off}}, \hat{d}_{\mathrm{off}} \right). \tag{10}$$

While the network may favor to optimize the class loss or the distance loss to reduce the total loss $E_{\mathrm{mt}}(\boldsymbol{\theta})$, the confidence loss encourages it to optimize both losses at the same time. This is expected to accelerate and facilitate the learning process.

### 2.5. Inference

Although an inference scheme similar to that in [10, 12] can be employed, we opted for a simple inference scheme here. Firstly, we performed thresholding on the posterior probability output by the background-rejection classifier with a threshold $\alpha_{\mathrm{prob}}$ to determine whether a sample should be classified as foreground and be directed to the event-classification classifier.

Moreover, we only made use of class labels obtained from the event-classification network, followed by median filtering with a window length $w_{\mathrm{sm}}$ for label smoothing. That is, we did not use the estimates for event onset and offset distances as in [10, 12]. This can be further explored in future work. Since three target event categories are evaluated separately in the challenge, when performing detection for a certain category, we ignored outputs of other categories. Lastly, non-maximum suppression was also applied. A maximum of one detected event with the longest duration was retained for each recording.

Table 3: Event-based overall performance of different systems on the development and test data.

| | Development data | | | | | | | | Evaluation data | | | |
|---|---|---|---|---|---|---|---|---|---|---|---|---|
| | DCASE baseline | | DNN | | CNN | | Best combination | | DCASE baseline | | Our submission | |
| | ER | F1 | ER | F1 | ER | F1 | ER | F1 | ER | F1 | ER | F1 |
| Baby cry | 0.67 | 72.0 | 0.36 | 80.5 | **0.09** | **95.3** | **0.09** | **95.3** | 0.80 | 66.8 | **0.23** | **88.4** |
| Glass break | 0.22 | 88.5 | **0.10** | **95.3** | 0.20 | 89.5 | **0.10** | **95.3** | 0.38 | 79.1 | **0.11** | **94.3** |
| Gun shot | 0.69 | 57.4 | **0.36** | **79.5** | 0.38 | 79.1 | **0.36** | **79.5** | 0.73 | 46.5 | **0.32** | **82.1** |
| Average | 0.53 | 72.7 | 0.27 | 85.1 | 0.22 | 88.0 | **0.18** | **90.0** | 0.64 | 64.1 | **0.22** | **88.2** |

## 3. EXPERIMENTS

### 3.1. DCASE 2017 development data

Our experiments were conducted on the development data of "Detection of rare events" task of the DCASE 2017 challenge [25]. Isolated events of three target categories: baby cry (106 training, 42 test instances), glass break (96 training, 43 test instances), and gun shot (134 training, 53 test instances) downloaded from freesound.org were mixed with background recordings from TUT Acoustic Scenes 2016 development dataset [26] to create 500 mixtures for each category in both training and test sets. The mixing event-to-background ratios (EBR) were -6, 0 and 6 dB. There are events in half of 500 mixtures, the other half is of only background. We made use of the standard data split provided by the challenge in the experiments.

### 3.2. Parameters

For the weighted loss in (2), we set $\lambda_{fg} = 10$ and $\lambda_{bg} = 1$. That is, false negatives are penalized ten times more than false positives. The associated weights of the multi-task loss in (5) were set to $\lambda_{class} = 1$, $\lambda_{dist} = 10$, and $\lambda_{conf} = 1$. We set $\lambda_{dist}$ larger than $\lambda_{class}$ and $\lambda_{conf}$ to encourage the networks to focus more on modeling event temporal structures. In addition, we set the regularization parameter $\lambda = 10^{-3}$ for both losses. The networks were trained using the *Adam* optimizer [27] with a learning rate of $10^{-4}$. The DNNs were trained for 200 epochs with a batch size of 256 whereas the CNNs were trained for 5 epochs with a batch size of 128.

In the inference step, the probability threshold $\alpha_{prob}$ was searched in the range of $[0, 1]$ with a step size of 0.05. In addition, we performed grid search for the smoothing window length $w_{sm}$ for each category in the range of $[3, 147]$ with a step size of 6. The values of $\alpha_{prob}$ and $w_{sm}$ yielding the best F-score were retained.

### 3.3. Experimental results on the development data

We used two event-based metrics for evaluation: detection error (ER) and F-score [28] as used for the challenge's baseline. We also compared the detection performances obtained by our systems to that of the DCASE 2017 baseline [25].

The detection performances obtained by different detection systems are shown in Table 3. As can be seen, the performances of the proposed DNN-based and CNN-based systems vary significantly for different event categories. While the former is more efficient in detecting glass break and gun shot events, the latter performs better on human-generated baby cry events. It seems that invariant features learned by a CNN, which are capable of handling the well-known vocal-tract length variation between speakers in speech recognition [29, 30, 31], are helpful for baby cry. In contrast, convolution does not help but worsens the detection performance of the non-human events (i.e. glass break and gun shot). Probably, these events do not possess the characteristics as human-generated events, and information in neighboring frequency bands should not be pooled. As a result, the DNN detector works better for these events than the CNN one, at least in our setup.

Both proposed DNN and CNN detectors significantly outperform the DCASE 2017 baseline over all three categories. On average, the DNN detector improves F-score to 85.1% from 72.7% of the baseline and reduces ER to 0.27 from 0.53 of the baseline. The CNN detector performs even better, achieving an F-score of 88.0% and an ER of 0.22. Our best combination system (i.e. the CNN system for baby cry and the DNN system for glass break and gun shot) achieves an F-score of 90.0% (i.e improving 17.3% absolute over that of the baseline) and an ER of 0.18 (i.e. reducing 0.35 absolute from that of the baseline).

## 4. THE SUBMISSION SYSTEM

Our submission system to Task 2 of the challenge is based on the best combination found in the experiments with the development data. That is, the CNN detector is in charge of detecting baby cry events while the DNN is responsible for detecting glass break and gun shot events. In combination with state-of-the-art phase-aware signal enhancement, the parameters that led to the best performance were retained to build the detection system, except for the smoothing window size $w_{sm}$. We experimentally saw a strong influence of this parameter on the detection performance of the development data. To avoid possible overfitting caused by this parameter, we chose the one that produces an event presence rate nearest to 0.5 which is the value used for generating the data [25]. The whole development data was used to train the detection system which was then tested on the challenge's evaluation data.

The results obtained by our submission system are shown in Table 3. Our system achives an F-score of 88.2% and an ER of 0.22 which are significantly better than those obtained by the DCASE baseline. Significant improvements on individual categories can also be seen. Note that we report the results here after correcting a minor mistake in our submission system. Therefore, they are slightly different from those reported in the official DCASE webpage, thanks to the organization team for re-evaluation. Overall, our team is ranked 3rd out of 13 participating teams.

## 5. CONCLUSIONS

We presented our proposed system participating in "Detection of rare events" task of the DCASE 2017 challenge. Two tailored loss functions were proposed to couple with DNNs and CNNs to address the common issues of audio event detection problem. The weighted loss is to tackle the data skewness issue in background/foreground classification and the multi-task loss enables the networks to jointly model event class distribution and event temporal structures for event classification. In combination with state-of-the-art phase-aware signal enhancement, we reported significant improvements in detection performance obtained by our proposed system over the challenge's baseline on both the development and evaluation data.